\documentclass[10pt]{iopartmod}
\usepackage{amssymb}
\usepackage{amsmath}
\usepackage{makeidx}
\usepackage[dvips]{graphicx}

\input{tcilatex}

\begin{document}

\title{Slow Wave Phenomena in Photonic Crystals}
\author{Alex Figotin and Ilya Vitebskiy}

\begin{abstract}
Slow light in photonic crystals and other periodic structures is associated
with stationary points of the photonic dispersion relation, where the group
velocity of light vanishes. We show that in certain cases, the vanishing
group velocity is accompanied by the so-called frozen mode regime, when the
incident light can be completely converted into the slow mode with huge
diverging amplitude. The frozen mode regime is a qualitatively new wave
phenomenon -- it does not reduce to any known electromagnetic resonance.
Formally, the frozen mode regime is not a resonance, in a sense that it is
not particularly sensitive to the size and shape of the photonic crystal.
The frozen mode regime is more robust and powerful, compared to any known
slow-wave resonance. It has much higher tolerance to absorption and
structural imperfections.

\end{abstract}

\maketitle

\section{Introduction}

\subsection{What is slow light?}

The velocity of light in transparent dispersive media is characterized by
two different physical quantities -- the phase velocity and the group
velocity. The phase velocity of a travelling wave is defined as%
\begin{equation}
v_{ph}=\omega /k,  \label{v_ph}
\end{equation}%
where $\omega $ is the frequency and $k$ is the wave number. The group
velocity of a travelling wave is defined as%
\begin{equation}
v_{g}=\partial \omega /\partial k.  \label{v_g}
\end{equation}%
The group velocity of light is one of the most important electromagnetic
characteristics of transparent media. It determines the speed of pulse
propagation and is usually referred to simply as the speed of light
propagation. With certain reservations, the group velocity also coincides
with electromagnetic energy velocity in a wave packet\cite{Brill,LLEM}. The
definitions (\ref{v_ph}) and (\ref{v_g}) apply not only to uniform
transparent media and waveguides, but also to nonuniform spatially periodic
structures, such as photonic crystals, arrays of coupled optical resonators,
etc. In the case of a nonuniform periodic structure, $k$ is the Bloch wave
number defined in the Brillouin zone \cite{Brill}. In a slow light case, the
electromagnetic pulse propagates through the dispersive medium with the
speed $v_{g}\ll c$, while the respective value of the phase velocity (\ref%
{v_ph}) is irrelevant.

The term "slow light" has been widely used to describe a broad range of
qualitatively different physical phenomena. There have been hundreds of
publications in which the above term appears as a key word, while in many
cases, the physical phenomena in question have little in common. Below, we
just briefly outline the most important situations in which the concept of
slow light can be invoked, and then we turn to the main subject of this
paper -- the frozen mode regime.

Most of the "slow light" cases can be grouped into two major categories:

\begin{itemize}
\item[-] those where the low group velocity of light results from strong
temporal (frequency) dispersion of the transparent medium;

\item[-] those where the low speed of pulse propagation is a result of
coherent interference in spatially periodic structures.
\end{itemize}

\subsection{Slow light in media with strong temporal dispersion}

In a uniform medium, both the phase and the group velocities of light can be
expressed in terms of the refractive index $n$%
\begin{equation}
v_{f}=c/n,\   \label{v_f (n)}
\end{equation}%
\begin{equation}
v_{g}=c\left( n+\omega \frac{dn}{d\omega }\right) ^{-1}.  \label{v_g (n)}
\end{equation}%
The derivative $dn/d\omega $ in (\ref{v_g (n)}) characterizes the frequency
dispersion of the medium. At optical frequencies, the refractive index $n$
of common transparent substances does not exceed several units, and the
speed of light propagation is of the same order of magnitude as in vacuum.
The situation can change dramatically in a strongly dispersive medium where
the term $\omega dn/d\omega $ in (\ref{v_g (n)}) becomes dominant%
\begin{equation}
\text{if }\omega \frac{dn}{d\omega }\gg n,\text{ then }\ v_{g}\approx
c\left( \omega \frac{dn}{d\omega }\right) ^{-1}\ll c.  \label{v_g<<c}
\end{equation}%
Strong frequency dependence of the refractive index $n$ implies that the
light group velocity can be substantially different from $c$. The relation (%
\ref{v_g<<c}) yields the following limitation on the slow pulse bandwidth 
\begin{equation}
\frac{\Delta \omega }{\omega }\ll \frac{v_{g}}{c},  \label{Delta om}
\end{equation}%
where the assumption is made that the refractive index $n$ within the
transparency window is of the order of unity. The condition (\ref{Delta om})
can also be viewed as a constraint on the minimal propagation speed of a
pulse with a given bandwidth $\Delta \omega $.

Usually, the large temporal dispersion (\ref{v_g<<c}) is a result of
excitation of electronic or some other intrinsic resonances of the medium.
This always involves some kind of a delicate resonant light-matter
interaction with extremely small bandwidth. For instance, in the well known
case of electromagnetically induced transparency \cite{EIT,EIT 1,EIT 2,EIT
3,EIT 4,EIT 5,EIT 6,EIT7}, the incident light interacts with atomic spin
excitations forming the so-called dark-state polaritons. These polaritons
propagate slowly through the medium in the form of a sharply compressed
pulse, the energy of which is much smaller than that of the incident light
pulse. In another example, the slow wave is associated with plasmons, which
are electron density waves in gas of mobile electrons of the metal. Plasmons
can also interact with light forming polaritons. There are many more
examples of condensed matter excitations, strongly interacting with light
and having low group velocity and relatively low relaxation rate. In most of
such cases, the slow pulse propagating through the dispersive medium can be
seen as an intrinsic coherent excitation (a polariton) triggered by the
input light, rather than a light pulse per se. The group velocity of such an
excitation may have little to do with the Maxwell equations and the speed of
light $c$.

In the rest of the paper we focus exclusively on the cases not involving any
intrinsic excitations of the medium and, therefore, not related to temporal
dispersion. Instead, we will focus on spatially periodic structures, in
which low speed of electromagnetic pulse propagation results solely from
spatial inhomogeniety.

\subsection{Slow light in spatially periodic structures}

Well-known examples of periodic dielectric structures include photonic
crystals \cite{PC Joann-95,PC Joann-03,PC-Scoda,PC Yariv}, periodic arrays
of coupled optical resonators \cite{CR B-02R,CR B-02,CR Meloni 03,CR Y-04,CR
Y-05,CR Kh-05,CR Kh-05OL,CR F-02,CR F-09}, and photonic crystal coupled
waveguide \cite{PCCW N-01,PCCW B-05,PCCW B-08,PCCW B-08R,PCCW B-07,PCCW
N-07,PCCW Ma-08,PCCW OE-07}. Low speed of pulse propagation in all these
cases is the result of coherent interference of light scattered at the
interfaces of adjacent structural components. The effects of spatial
dispersion associated with structural periodicity are particularly strong
when the structural period $L$ and the light wavelength $\lambda $ are
comparable in value%
\begin{equation}
L\sim \lambda .  \label{L=lambda}
\end{equation}

Strong spatial dispersion can result in low or even zero group velocity of
the respective Bloch wave. In practice, at optical frequencies, the speed of
pulse propagation in periodic dielectric arrays \ can be reduced by not more
than three orders of magnitude. This is not a fundamental restriction, but
rather a technological limitation related to the difficulty of building
flawless periodic arrays at nanoscales. On the positive side, the dielectric
components of the periodic array are not required to display strong temporal
dispersion. As a consequence, the absorption of light is not an inherent
problem in this case.

There is a natural bandwidth limitation on the slow pulse in periodic
dielectric arrays, which is similar to the case of slow light in
time-dispersive media. Let $\Delta \omega $ be the frequency bandwidth of a
pulse and $\Delta k$ -- the respective range of the Bloch wave number. The
average group velocity $\left\langle v_{g}\right\rangle $ of the pulse can
be defined as%
\begin{equation}
\left\langle v_{g}\right\rangle \approx \frac{\Delta \omega }{\Delta k}.
\label{<v_g>}
\end{equation}%
Let us assume that the pulse propagating in the periodic medium is composed
of the Bloch eigenmodes belonging to the same spectral branch of the
dispersion relation $\omega \left( k\right) $. This assumption implies that $%
\Delta k$ cannot exceed the size $2\pi /L$ of the Brillouin zone%
\begin{equation*}
\Delta k<2\pi /L,
\end{equation*}%
where $L$ is the unit cell length of the periodic array. In addition, we
assume that the refractive index of the constitutive components of the
periodic array is of the order of unity and, therefore,%
\begin{equation}
L\sim \lambda _{0}=2\pi c/\omega ,  \label{L=c/om}
\end{equation}%
where $\lambda _{0}$ is the light wavelength in vacuum. The relations (\ref%
{<v_g>}) and (\ref{L=c/om}) yield the following limitation on the minimal
propagation speed of a pulse with a given bandwidth $\Delta \omega $%
\begin{equation}
\left\langle v_{g}\right\rangle >\frac{L}{2\pi }\Delta \omega \sim c\frac{%
\Delta \omega }{\omega }.  \label{v_g min}
\end{equation}%
The restriction (\ref{v_g min}) is similar to that defined by the inequality
(\ref{Delta om}) and related to the case of slow light in a uniform medium
with temporal (frequency) dispersion.

\subsection{Examples of periodic arrays supporting slow light}

During the last two decades, a tremendous progress has been made in theory
and applications of periodic arrays of coupled optical resonators.
Generally, if the coupling between adjacent resonators in a periodic array
is weak, the group velocity of Bloch excitations propagating through the
array is very low in each and every transmission band. This is true
regardless of the nature of individual resonators. The above simple idea
forms the basis for one of the most popular approaches to slowing down the
light. One inevitable consequence of the weak coupling between the
neighboring optical resonators is that all the individual transmission bands
are very narrow. Another problem is that there can be severe restrictions on
transmitted power. An extensive discussion on the subject and numerous
examples and references can be found in \cite{CR B-02R,CR B-02,CR Meloni
03,CR Y-04,CR Y-05,CR Kh-05,CR Kh-05OL,CR F-02,CR F-09}. A qualitatively
similar situation occurs in line-defect waveguides in photonic crystals,
where a periodic array of structural defects plays the role of weakly
coupled optical resonators \cite{PCCW N-01,PCCW B-05,PCCW B-08,PCCW
B-08R,PCCW B-07,PCCW N-07,PCCW Ma-08,PCCW OE-07}.

Slow light phenomena in periodic arrays of weakly coupled resonators have
been the subject of a great number of recent publications, including some
excellent review articles cited above. A common characteristic of all
different realizations of this approach is a relatively low density of
modes. One consequence of this is a significant nonlinearity. In certain
cases, the nonlinearity is so extreme that it occurs on a single photon
level. Strong nonlinearity can be useful in controlling the flow of light.
On the other hand, it can severely limit the transmission capabilities of
the optical waveguide. In this respect, photonic crystals have advantage.

Photonic crystals are spatially periodic structures composed of two or more
different transparent dielectric materials \cite{PC Joann-95,PC
Joann-03,PC-Scoda,PC Yariv}. Unlike the case of optical waveguides and
linear arrays of coupled resonators, in photonic crystals we have bulk
electromagnetic waves capable of propagating in any direction through the
periodic heterogeneous structure. This results in much greater density of
modes, compared to arrays of weakly coupled resonators \cite{SL Scal1,SL
Scal2,SL Joann}.

Strong spatial dispersion of a typical photonic crystal and any other
periodic dielectric structure is reflected in a complicated $k-\omega $
diagram featuring transmission bands and gaps. Normally, each spectral
branch $\omega \left( k\right) $ of the Bloch dispersion relation develops
stationary points $\omega _{s}=\omega \left( k_{s}\right) $ where the group
velocity (\ref{v_g}) of the corresponding propagating mode vanishes%
\begin{equation}
d\omega /dk=0\text{, at }\omega =\omega _{s}=\omega \left( k_{s}\right) .%
\text{ }  \label{SP}
\end{equation}%
Examples of different stationary points are shown in Fig. \ref{DRSP3}. Each
of the frequencies $\omega _{g}$, $\omega _{0}$ and $\omega _{d}$ is
associated with zero group velocity of the respective traveling wave, but
there are some fundamental differences between those three cases. These
differences are particularly pronounced when it comes to the efficiency of
conversion of the incident light into the slow mode inside the periodic
structure. In most cases, incident radiation with the frequency of one of
the slow modes is simply reflected back to space, without exciting the slow
mode inside the periodic structure. How to overcome this fundamental problem
and, thereby, how to transform a significant fraction of the incident light
energy into a slow mode with drastically enhanced amplitude, is the primary
subject of this paper.

\section{Frozen mode regime in photonic crystals.}

The effects of strong spatial dispersion in periodic dielectric structures
culminate at stationary points (\ref{SP}) of the Bloch dispersion relation,
where the group velocity (\ref{v_g}) of a traveling Bloch wave vanishes. One
reason for this is that vanishing group velocity always implies a dramatic
increase in density of modes at the respective frequency. In addition,
vanishing group velocity also implies certain qualitative changes in the
eigenmode structure, which can strongly affect the propagation and
scattering of light. A particular example of the kind is the frozen mode
regime associated with a dramatic enhancement of the amplitude of the wave
transmitted to the periodic medium \cite{PRB03,PRE03,PRE05B,PRE06}. There
are at least two qualitatively different modifications of the frozen mode
regime, each related to a specific singularity of the electromagnetic
dispersion relation. Both effects can be explained using a simple example of
a plane electromagnetic wave normally incident on a lossless semi-infinite
periodic structure.

\begin{figure}[tbph]
\scalebox{0.8}{\includegraphics[viewport=0 0 500 180,clip]{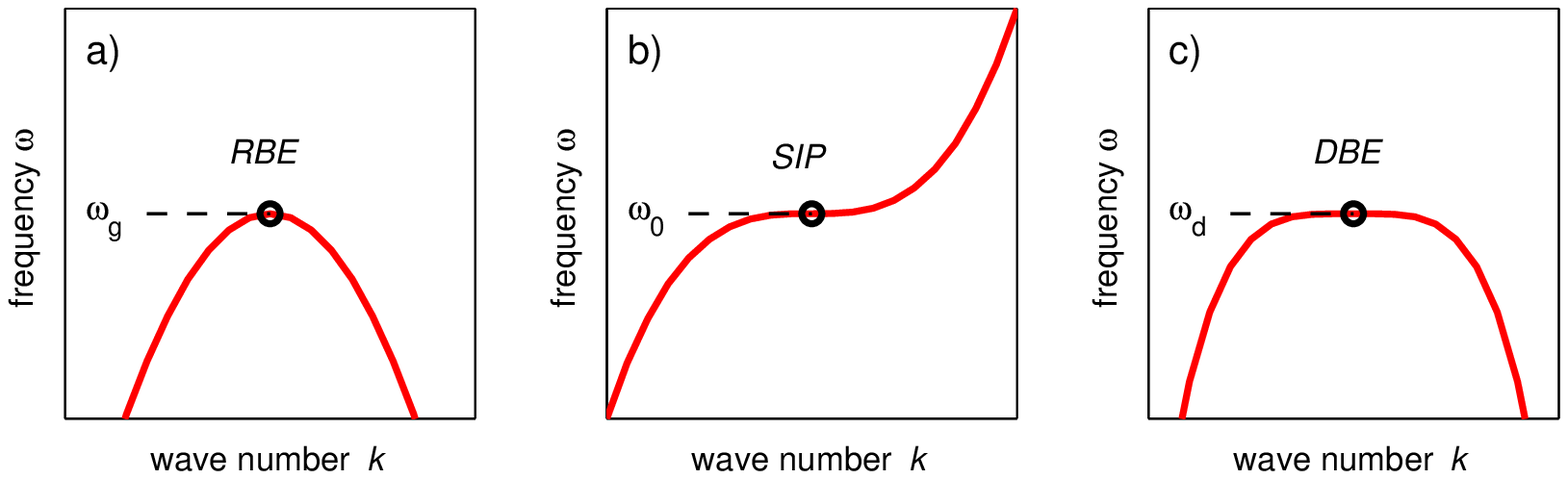}}
\caption{(Color online). Schematic examples of dispersion relations
displaying different stationary points: (a) a regular band edge (RBE), (b) a
stationary inflection point (SIP), (c) a degenerate band edge (DBE).}
\label{DRSP3}
\end{figure}

\begin{figure}[tbph]
\scalebox{0.8}{\includegraphics[viewport=0 0 500 450,clip]{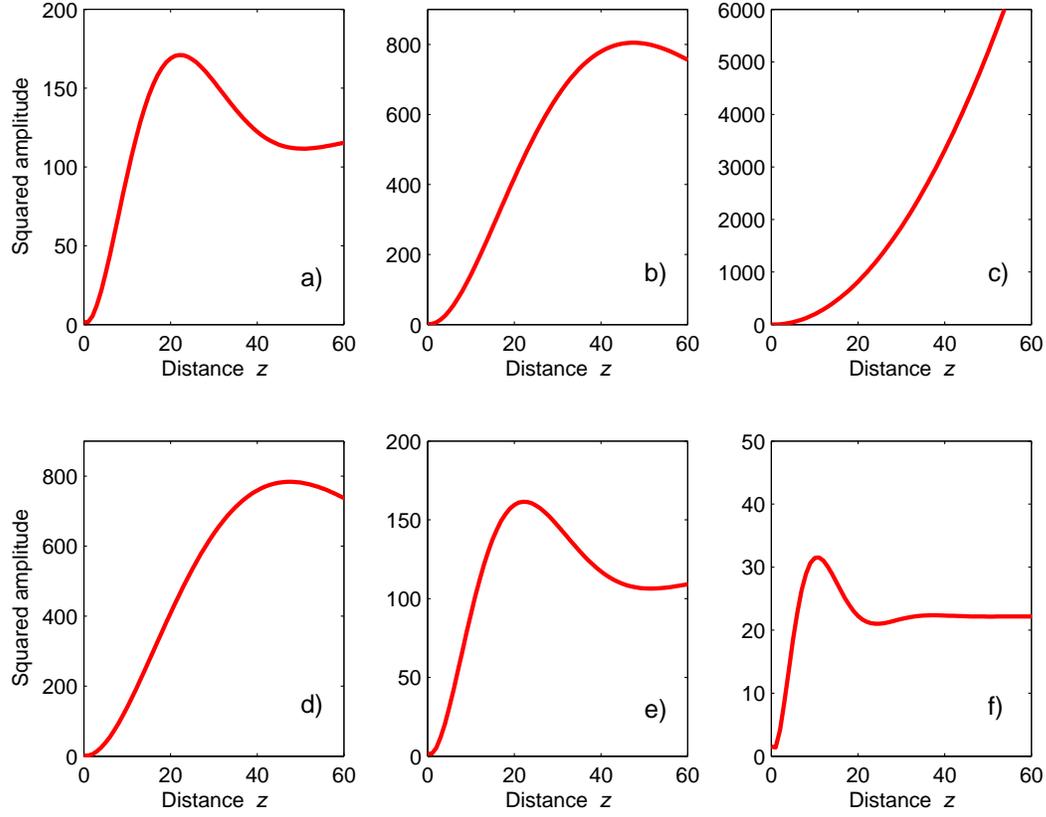}}
\caption{(Color online). Smoothed profile of the frozen mode at six
different frequencies in the vicinity of stationary inflection point: (a) $%
\protect\omega =\protect\omega _{0}-10^{-4}c/L$, (b) $\protect\omega =%
\protect\omega _{0}-10^{-5}c/L$, (c) $\protect\omega =\protect\omega _{0}$,
(d) $\protect\omega =\protect\omega _{0}+10^{-5}c/L$, (e) $\protect\omega =%
\protect\omega _{0}+10^{-4}c/L$, (f) $\protect\omega =\protect\omega %
_{0}+10^{-3}c/L$. In all cases, the incident wave has the same polarization
and unity amplitude. The distance $z$ from the surface of semi-infinite
photonic crystal is expressed in units of $L$.}
\label{AMn6w0}
\end{figure}

The frozen mode regime of the first kind is associated with a stationary
inflection point (SIP) of the $k-\omega $ diagram shown in Fig. \ref{DRSP3}%
(b). In the vicinity of stationary inflection point, the relation between
the frequency $\omega $\ and the Bloch wave number $k$ can be approximated as%
\begin{equation}
\omega -\omega _{0}\propto \left( k-k_{0}\right) ^{3}.  \label{SIP DR}
\end{equation}%
A monochromatic plane wave of frequency close to $\omega _{0}$ incident on
semi-infinite photonic crystal is converted into the frozen mode with
infinitesimal group velocity and dramatically enhanced amplitude, as
illustrated in Fig. \ref{AMn6w0}. The saturation value of the frozen mode
amplitude diverges as the frequency approaches the SIP value $\omega _{0},$
as shown in Fig. \ref{AMn6w0}(c). Remarkably, the photonic crystal
reflectivity at $\omega =\omega _{0}$ can be very low, implying that the
incident radiation is almost totally converted into the frozen mode with
zero group velocity, drastically enhanced amplitude, and finite energy flux
close to that of the incident wave. This remarkable phenomenon is uniquely
associated with a stationary inflection point (SIP) of the Bloch dispersion
relation. Not every periodic structure can display such a spectral
singularity, but if it does, one can expect the behavior described in Fig. %
\ref{AMn6w0}.

\begin{figure}[tbph]
\scalebox{0.8}{\includegraphics[viewport=0 0 500 450,clip]{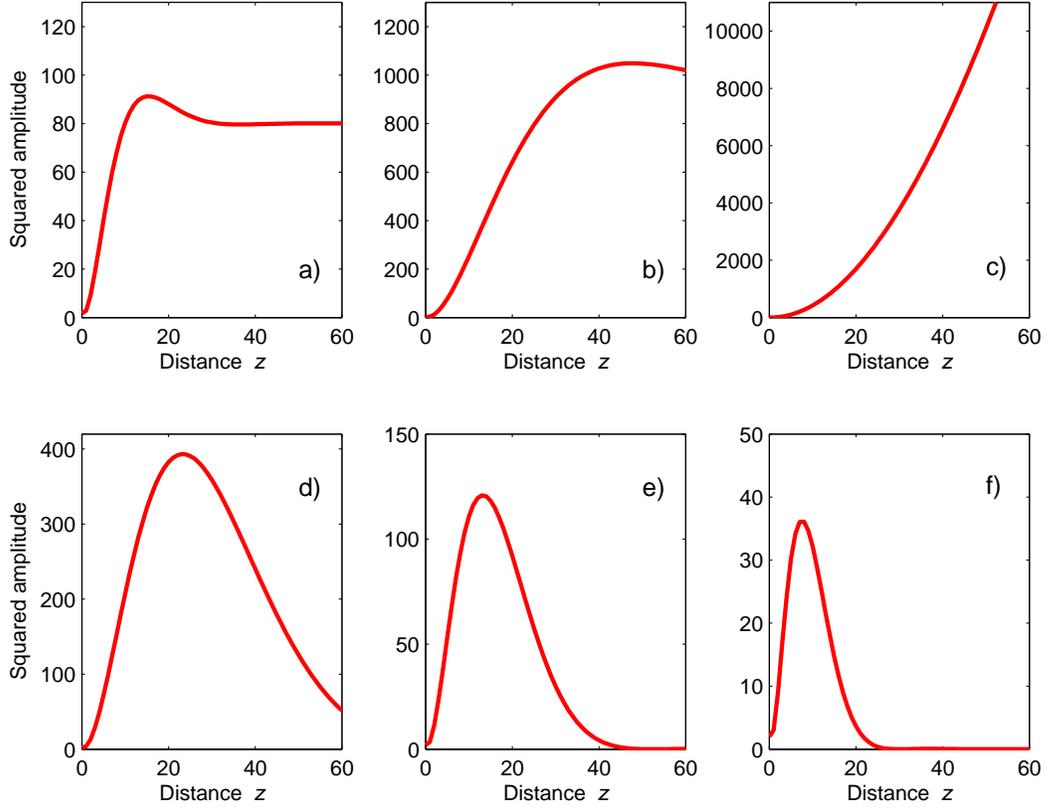}}
\caption{(Color online). Smoothed profile of the frozen mode at six
different frequencies in the vicinity of degenerate band edge: (a) $\protect%
\omega =\protect\omega _{d}-10^{-4}c/L$, (b) $\protect\omega =\protect\omega %
_{d}-10^{-6}c/L$, (c) $\protect\omega =\protect\omega _{d}$, (d) $\protect%
\omega =\protect\omega _{d}+10^{-6}c/L$, (e) $\protect\omega =\protect\omega %
_{d}+10^{-5}c/L$, (f) $\protect\omega =\protect\omega _{d}+10^{-4}c/L$. In
the transmission band (at $\protect\omega <\protect\omega _{d}$), the
asymptotic field value diverges as $\protect\omega \rightarrow \protect%
\omega _{d}$. By contrast, in the band gap (at $\protect\omega >\protect%
\omega _{d}$), the asymptotic field value is zero. The amplitude of the
incident wave at $z<0$ is unity. The distance $z$ from the surface is
expressed in units of $L$.}
\label{Amn6wd}
\end{figure}

A qualitatively different kind of frozen mode regime can occur in the
vicinity of a degenerate photonic band edge (DBE) shown in Fig. \ref{DRSP3}%
(c). At frequencies just below $\omega _{d}$, the respective dispersion
relation can be approximated as%
\begin{equation}
\omega _{d}-\omega \propto \left( k-k_{d}\right) ^{4},\text{ at }\omega
\lessapprox \omega _{d}.  \label{DBE DR}
\end{equation}%
Contrary to the case of stationary inflection point (\ref{SIP DR}), in the
vicinity of a degenerate band edge the photonic crystal becomes totally
reflective. But at the same time, the steady-state field inside the periodic
medium (at $z>0$) develops a very large amplitude, diverging as the
frequency approaches its critical value $\omega _{d}$. Such a phenomenon is
illustrated in Fig. \ref{Amn6wd}. The frozen mode profile below and above
the degenerate band edge frequency $\omega _{d}$ is different. It has a
large saturation value at frequencies located inside the transmission band
(at $\omega \leq \omega _{d}$), as seen in Fig. \ref{Amn6wd}(a) and (b).
This saturation value diverges as $\omega \rightarrow \omega _{d}-0$. By
contrast, at frequencies lying inside the band gap (at $\omega >\omega _{d}$%
), the field amplitude initially increases dramatically with the distance $z$
from the surface, but then vanishes as the distance $z$ further increases,
as seen in Fig. \ref{Amn6wd}(d -- f).

Hereinafter, the above two phenomena will be referred to as the SIP-related
frozen mode regime and the DBE-related frozen mode regime, respectively.
Figs. \ref{AMn6w0} and \ref{Amn6wd} describe the frozen mode regime in
hypothetical lossless semi-infinite photonic crystal. What happens to the
frozen mode regime in a plane-parallel photonic slab with a finite thickness 
$D$?

\begin{figure}[tbph]
\scalebox{0.8}{\includegraphics[viewport=0 0 500 450,clip]{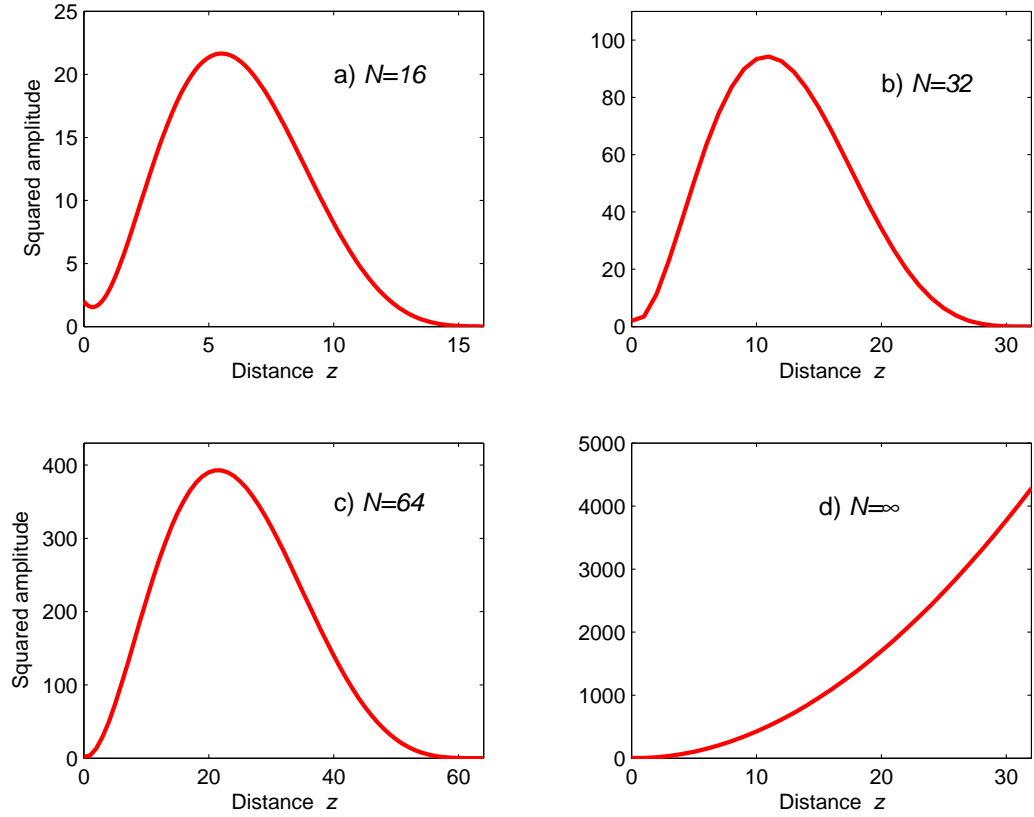}}
\caption{(Color online). Smoothed profile of the frozen mode in periodic
layered structures composed of different number $N$ of unit cells $L$. The
frequency is equal to that of the degenerate band edge. The initial rate of
growth of the frozen mode amplitude is virtually independent of $N$. The
limiting case (d) of the semi-infinite structure is also shown in Fig. 
\protect\ref{Amn6wd}(c). In all cases, the incident wave has the same
polarization and unity amplitude. The distance $z$ from the surface is
expressed in units of $L$.}
\label{SMNnd}
\end{figure}

First, let us assume that the incident light frequency is equal to that of
the respective stationary point ($\omega _{0}$ or $\omega _{d}$) of the
Bloch dispersion relation in Fig. \ref{DRSP3}. In either case, in the
leftmost portion of the photonic slab, the frozen mode profile remains the
same as in the semi-infinite case shown in Figs. \ref{AMn6w0}(c) and \ref%
{Amn6wd}(c). In the opposite, rightmost part of the photonic slab, the
frozen mode amplitude now vanishes, as shown in Fig. \ref{SMNnd}(a-c). In
either case, the maximum field intensity $W_{\max }$ inside the photonic
slab is%
\begin{equation}
W_{\max }\propto W_{I}N^{2},  \label{W - N2}
\end{equation}%
where $W_{I}$ is the intensity of the incident light and $N=D/L$ is the
number of periods (primitive translations) in the $z$ direction of the
finite periodic structure. Additional factors limiting the frozen mode
amplitude include structural imperfections of the periodic array,
absorption, nonlinearity, deviation of the incident radiation from plane
monochromatic wave, etc.

Now, what happens in a finite photonic slab if the incident light frequency
is slightly different from that of the respective stationary point $\omega
_{0}$ or $\omega _{d}$?

In the case of a SIP-related frozen mode regime, the field amplitude will
decrease as the frequency $\omega $ deviates from $\omega _{0}$ in either
direction -- similar to the case of semi-infinite structure shown in Fig. %
\ref{AMn6w0}. In other words, in a lossless finite photonic slab, the
SIP-related frozen mode regime appears as a resonance centered at $\omega
_{0}$. According to (\ref{W - N2}), the light intensity at the resonance is
proportional to $N^{2}$, while the respective Q-factor is proportional to $%
N^{3}$.

\begin{figure}[tbph]
\scalebox{0.8}{\includegraphics[viewport=0 0 500 450,clip]{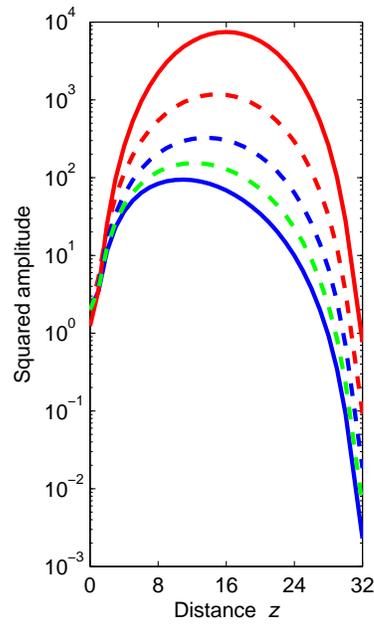}}
\caption{(Color online) Steady-state field distribution in a lossless
periodic finite slab with $N=32$ at a series of equidistant frequencies
lying between the DBE frequency $\protect\omega _{d}$ and the frequency $%
\protect\omega _{r}$ of the giant slow wave resonance. Observe that the
field amplitude increases dramatically as the frequency changes from that of
the DBE (the lowest solid blue curve) to that of the giant slow wave
resonance (the upper solid red curve). The lowest solid curve is identical
to that in Fig. \protect\ref{SMNnd}(b).}
\label{SM32g0w}
\end{figure}

In the case of a DBE-related frozen mode regime, the situation is quite
different. Namely, in a lossless, plane-parallel photonic sample with a
finite thickness, the DBE-related frozen mode regime is overwhelmed with
much more powerful, giant slow wave resonance \cite{PRE05,PRA07}. Let us
explain this phenomenon in more detail. At DBE frequency $\omega _{d}$, the
frozen mode profile in a finite photonic slab is shown in Fig. \ref{SMNnd}%
(a-c). According to (\ref{W - N2}), the maximum light intensity inside the
photonic slab is proportional to $N^{2}$. If the incident light frequency
deviates from $\omega _{d}$ toward the photonic band gap in Fig. \ref{DRSP3}%
(c), the light intensity inside the photonic slab will decrease, similar to
the semi-infinite case shown in Figs. \ref{Amn6wd}(d-f). But if the incident
light frequency deviates from $\omega _{d}$ in the opposite direction
(toward the transmission band in Fig. \ref{DRSP3}(c)), the light intensity
inside the photonic slab will increase dramatically, peaking at a certain
resonance frequency $\omega _{r}$, as shown in Fig. \ref{SM32g0w}. The exact
location of the resonance frequency $\omega _{r}$ is dependent on the
thickness of the finite periodic structure \cite{PRE05,PRA07}. But in any
event, $\omega _{r}$ lies inside the transmission band and approaches the
DBE frequency $\omega _{d}$ as $N$ increases. The light intensity at the
frequency $\omega _{r}$ of the giant slow wave resonance is%
\begin{equation}
W_{\max }\propto W_{I}N^{4},  \label{W - N4}
\end{equation}%
which is by factor $N^{2}$ larger than that of the common slow wave
resonance associated with the regular photonic band edge in Fig. \ref{DRSP3}%
(a). The Q-factor of the giant slow wave resonance is proportional to $N^{5}$%
.

In summary, in a photonic slab of finite thickness, the SIP-related frozen
mode regime persists, but its amplitude is limited by the slab thickness in
accordance with the relation (\ref{W - N2}). By contrast, the DBE-related
frozen mode regime in a finite plane-parallel photonic slab is overrun by
the giant slow wave resonance, which can be viewed as a combination of the
frozen mode regime and the Fabry-Perot resonance. It is much more powerful
compared to the common RBE-related slow wave resonance in a photonic slab of
similar size \cite{PRE05,PRA07}.

Not every periodic structure can support the frozen mode regime at normal
incidence. Generally, the physical conditions for the frozen mode regime are
the same as the conditions for the existence of the respective stationary
points (\ref{SIP DR}) or (\ref{DBE DR}) of the Bloch dispersion relation.
For instance, in the case of periodic layered structure, a unit cell must
contain at least three layers, of which two must display a misaligned
in-plane anisotropy \cite{PRE01,PRB03,PRE06}. At optical frequencies,
fabrication of such a layered structure can be challenging. Fortunately, in
photonic crystals with $3D$ periodicity, the conditions for the existence of
stationary inflection point (SIP) and the related frozen mode regime are
much less restrictive. A good example of the kind is provided by the inverse
opal structure, where such a spectral singularity was found and
investigated, both theoretically and experimentally in \cite{Hui08}.

The frozen mode regime is a qualitatively new wave phenomenon -- it does not
reduce to any known electromagnetic resonance. Formally, the SIP-related
frozen mode regime is not a resonance, in a sense that it is not
particularly sensitive to the size and shape of the photonic crystal.
Besides, the frequency dependence of the frozen mode amplitude is very
different from that of a cavity resonance, or a common slow wave resonance.
The frozen mode regime is much more robust, compared to any known slow-wave
resonance occurring in periodic and nonperiodic photonic structures. It has
much higher tolerance to absorption and structural imperfections than common
Fabry-Perot or transmission band edge resonances, where the entire photonic
crystal works as a resonator.

The DBE-related frozen mode regime is quite different. In a finite
plane-parallel slab it is overrun with the much more powerful giant
slow-wave resonance \cite{PRE05,PRA07}. The Q-factor associated with such a
resonance can be by two orders of magnitude higher, compared to that of the
regular slow-wave resonance in the same or similar periodic structure. More
importantly, it provides the possibility of a dramatic reduction in size --
up to an order of magnitude -- of some basic photonic devices without
compromising their performance.

\section{Physical nature of the frozen mode regime}

The essence of the frozen mode regime can be understood from the simple
example of a plane monochromatic wave normally incident on a semi-infinite
periodic layered structure. An important requirement, though, is that some
of the layers display a misaligned in-plane anisotropy \cite%
{PRE01,PRB03,PRE06}. Below we present a comparative analysis of two
different kinds of frozen mode regime. Although throughout this section we
only consider the case of normal incidence, most of the results and
expressions remain virtually unchanged in a more general case of the frozen
mode regime at oblique propagation \cite{PRE03,PRE05B,PRE06}. One
difference, though, is that at oblique incidence, the frozen mode regime can
occur in much simpler structures, which can have a big practical advantage.

To start with, let us introduce some basic notations and definitions. Let $%
\Psi _{I}$, $\Psi _{R}$, and $\Psi _{T}$ be the incident, reflected and
transmitted waves, respectively. We assume that all three monochromatic
waves propagate along the $z$ axis normal to the layers. Electromagnetic
field both inside (at $z>0$) and outside (at $z<0$) the semi-infinite
periodic structure is independent of the $x$ and $y$ coordinates. The
transverse field components can be represented as a column-vector 
\begin{equation}
\Psi \left( z\right) =\left[ 
\begin{array}{c}
E_{x}\left( z\right) \\ 
E_{y}\left( z\right) \\ 
H_{x}\left( z\right) \\ 
H_{y}\left( z\right)%
\end{array}%
\right] ,  \label{Psi}
\end{equation}%
where $\vec{E}\left( z\right) $ and $\vec{H}\left( z\right) $ are
time-harmonic electric and magnetic fields. All four transverse field
components in (\ref{Psi}) are continuous functions of $z$, which leads to
the following standard boundary condition at the photonic crystal interface
at $z=0$ 
\begin{equation}
\Psi _{T}\left( 0\right) =\Psi _{I}\left( 0\right) +\Psi _{R}\left( 0\right)
.  \label{BC}
\end{equation}%
Assume also that anisotropic layers of the periodic array have an in-plane
birefringence%
\begin{equation*}
\left[ 
\begin{array}{ccc}
\varepsilon _{xx} & \varepsilon _{xy} & 0 \\ 
\varepsilon _{xy}^{\ast } & \varepsilon _{yy} & 0 \\ 
0 & 0 & \varepsilon _{zz}%
\end{array}%
\right] ,
\end{equation*}%
in which case the fields $\vec{E}\left( z\right) $ and $\vec{H}\left(
z\right) $ are normal to the direction of light propagation%
\begin{equation}
\vec{E}\left( z\right) \perp z,\vec{H}\left( z\right) \perp z.
\label{Ez=Hz=0}
\end{equation}%
Note that the polarizations of the incident, reflected, and transmitted
waves can be all different, because some of the layers of the periodic array
are birefringent. The presence of birefringent layers is essential for the
possibility of frozen mode regime. Moreover, in the case of normal
incidence, each unit cell of the periodic stack\ must include at least two
birefringent layers with misaligned anisotropy axes \cite{PRE01,PRB03,PRE06}.

In periodic layered media, the electromagnetic eigenmodes $\Psi _{k}\left(
z\right) $ are usually chosen in the Bloch form%
\begin{equation}
\Psi _{k}\left( z+L\right) =e^{ikL}\Psi _{k}\left( z\right) ,  \label{BF}
\end{equation}%
where the Bloch wave number $k$ is defined up to a multiple of $2\pi /L$.
The correspondence between $\omega $ and $k$ is referred to as the Bloch
dispersion relation. Real wave numbers $k$ correspond to propagating
(traveling) Bloch modes. Propagating modes belong to different spectral
branches $\omega \left( k\right) $ separated by frequency gaps. The speed of
a traveling wave in a periodic medium is determined by the group velocity (%
\ref{v_g}). Complex wave numbers $k=k^{\prime }+ik^{\prime \prime }$
correspond to evanescent Bloch modes. Evanescent modes decay exponentially
with the distance $z$ from the boundary of semi-infinite periodic structure.
Therefore, under normal circumstances, evanescent contribution to the
transmitted wave $\Psi _{T}\left( z\right) $ can be significant only in
close proximity of the photonic crystal interface at $z=0$. The situation
can change dramatically when the frequency $\omega $ approaches one of the
stationary point values $\omega _{s}$. At first sight, stationary points (%
\ref{SP}) relate only to propagating Bloch modes. But in fact, in the
vicinity of every stationary point frequency $\omega _{s}$, the imaginary
part $k^{\prime \prime }$ of the Bloch wave number of at least one of the
evanescent modes also vanishes. As a consequence, the respective evanescent
mode decays very slowly, and its role may extend far beyond the photonic
crystal boundary. In addition, in the special cases of interest, the
electromagnetic field distribution $\Psi \left( z\right) $ in the coexisting
evanescent and propagating eigenmodes becomes very similar, as $\omega $
approaches $\omega _{s}$. This can result in an abnormal interference
pattern constituting the frozen mode regime. What exactly happens in the
vicinity of a particular stationary point (\ref{SP}) essentially depends on
its character and appears to be very different in each of the three cases
presented in Fig. \ref{DRSP3}.

\subsection{Bloch composition of frozen mode}

If all the layers of the periodic layered structure were made of isotropic
materials, such as glass or air, each of the Bloch eigenmodes would be
doubly degenerate with respect to light polarization. This applies both to
propagating and evanescent Bloch modes. In our case, though, each unit cell
of the periodic array contains a pair of birefringent layers with misaligned
anisotropy axes. As a consequence, the Bloch eigenmodes with different
(elliptic) polarization will not be degenerate. At any given frequency, the
total number of Bloch eigenmodes with $k\parallel z$ is four. The only
exception are the stationary point frequencies (\ref{SP}), where some of the
four eigenmodes cannot be represented in the Bloch form (\ref{BF}). Those
cases will be discussed later.

With the exception of stationary points (\ref{SP}), any plane monochromatic
wave (\ref{Psi}) can be represented as a superposition of four Bloch
eigenmodes, propagating and/or evanescent, with different polarizations and
wave numbers. But in the setting where the semi-infinite periodic array
occupies the half-space $z\geq 0$, the transmitted wave is a superposition
of only two of the four Bloch eigenmodes. Indeed, neither the propagating
modes with negative group velocity, nor evanescent modes exponentially
growing with the distance $z$ from the surface, contribute to $\Psi
_{T}\left( z\right) $ in this case. Generally, one can distinguish three
different possibilities.

\begin{enumerate}
\item Both Bloch components of the transmitted wave $\Psi _{T}$ are
propagating modes%
\begin{equation}
\Psi _{T}\left( z\right) =\Psi _{pr1}\left( z\right) +\Psi _{pr2}\left(
z\right) ,\ \;z\geq 0.  \label{PsiT=pr+pr}
\end{equation}%
$\Psi _{pr1}\left( z\right) $ and $\Psi _{pr2}\left( z\right) $ are two
propagating Bloch modes with different real wave numbers $k_{1}$ and $k_{2}$
and different group velocities $v_{g1}>0$ and $v_{g2}>0$. This constitutes
the phenomenon of double refraction, provided that $v_{g1}$ and $v_{g2}$ are
different. The remaining two Bloch modes of the same frequency have negative
group velocities and cannot contribute to the transmitted wave $\Psi _{T}$.

\item Both Bloch components of $\Psi _{T}$ are evanescent%
\begin{equation}
\Psi _{T}\left( z\right) =\Psi _{ev1}\left( z\right) +\Psi _{ev2}\left(
z\right) ,\ \;z\geq 0.  \label{PsiT=ev+ev}
\end{equation}%
The respective two values of $k$ are complex with positive imaginary parts $%
k_{1}^{\prime \prime }>0,$ $k_{2}^{\prime \prime }>0$. This is the case when
the frequency $\omega $ falls into photonic band gap at $\omega >\omega _{g}$
in Fig. \ref{DRSP3}(a) or at $\omega >\omega _{d}$ in Fig. \ref{DRSP3}(c).
The fact that $k_{1}^{\prime \prime }>0,$ $k_{2}^{\prime \prime }>0$ implies
that the wave amplitude decays with the distance $z$ from the surface. In
the case (\ref{PsiT=ev+ev}), the incident wave is totally reflected back to
space by the semi-infinite periodic structure.

\item One of the Bloch components of the transmitted wave $\Psi _{T}$ is a
propagating mode with $v_{g}>0$, while the other is an evanescent mode with $%
k^{\prime \prime }>0$%
\begin{equation}
\Psi _{T}\left( z\right) =\Psi _{pr}\left( z\right) +\Psi _{ev}\left(
z\right) ,\ \;z\geq 0.  \label{PsiT=pr+ev}
\end{equation}%
For example, this is the case at $\omega \sim \omega _{0}$ in Fig. \ref%
{DRSP3}(b), as well as at $\omega <\omega _{g}$ in Fig. \ref{DRSP3}(a) and
at $\omega <\omega _{d}$ in Fig. \ref{DRSP3}(c). As the distance $z$ from
the surface increases, the evanescent contribution $\Psi _{ev}$ in (\ref%
{PsiT=pr+ev}) decays as $\exp \left( -zk^{\prime \prime }\right) $, and the
resulting transmitted wave $\Psi _{T}\left( z\right) $ turns into a single
propagating Bloch mode $\Psi _{pr}$.
\end{enumerate}

Propagating modes with $v_{g}>0$ and evanescent modes with $k^{\prime \prime
}>0$ are referred to as \emph{forward} modes. Only forward modes contribute
to the transmitted wave $\Psi _{T}\left( z\right) $ in the case of a
periodic semi-infinite stack. The propagating modes with $v_{g}<0$ and
evanescent modes with $k^{\prime \prime }<0$ are referred to as \emph{%
backward} modes. The backward Bloch modes never contribute to the
transmitted wave $\Psi _{T}\left( z\right) $ inside the periodic
semi-infinite stack. This statement is based on the following two
assumptions:

\begin{itemize}
\item[-] The transmitted wave $\Psi _{T}$ and the reflected wave $\Psi _{R}$
are originated from the plane wave $\Psi _{I}$ incident on the semi-infinite
photonic slab from the left.

\item[-] The periodic structure occupies the entire half-space and is
perfectly periodic at $z>0$.
\end{itemize}

If either of the above conditions is violated, the electromagnetic field
inside the periodic stack can be a superposition of four Bloch eigenmodes
with either sign of the group velocity $v_{g}$ of propagating contributions,
or either sign of $k^{\prime \prime }$\ of evanescent contributions. This
would be the case if the periodic array had some kind of structural defects
or a finite thickness. At the end of this section we briefly discuss how it
would affect the frozen mode regime.

Note also that the assumption that the transmitted wave $\Psi _{T}\left(
z\right) $ is a superposition of propagating and/or evanescent Bloch
eigenmodes may not apply if the frequency $\omega $ exactly coincides with
one of the stationary point frequencies (\ref{SP}). For example, at
frequency $\omega _{0}$ of stationary inflection point (\ref{SIP DR}), there
are no evanescent solutions to the Maxwell equations, and the transmitted
wave $\Psi _{T}\left( z\right) $ is a (non-Bloch) Floquet eigenmode linearly
growing with $z$ \cite{PRB03,PRE03}. Similar situation occurs at frequency $%
\omega _{d}$ of degenerate band edge (\ref{DBE DR}). The term "non-Bloch"
means that the respective field distribution does not comply with the
relation (\ref{BF}). At the same time, at any general frequency, including
the vicinity of any stationary point (\ref{SP}), the transmitted wave $\Psi
_{T}\left( z\right) $ is a superposition of two forward Bloch eigenmodes,
each of which is either propagating, or evanescent.

In all three cases (\ref{PsiT=pr+pr} -- \ref{PsiT=pr+ev}), the contribution
of a particular Bloch eigenmode to the transmitted wave $\Psi _{T}$ depends
on the polarization $\Psi _{I}$ of the incident wave. One can always choose
some special (elliptic) incident wave polarization, such that only one of
the two forward Bloch modes is excited and the transmitted wave $\Psi _{T}$
is a single Bloch eigenmode. In the next subsection we will see that there
is no frozen mode regime in the case of a single mode excitation. This fact
relates to the very nature of the frozen mode regime in a periodic layered
structure \cite{PRE06}.

Knowing the Bloch composition of the transmitted wave we can give a
semi-qualitative description of what happens when the frequency $\omega $ of
the incident wave approaches one of the stationary points (\ref{SP}) in Fig. %
\ref{DRSP3}. The rigorous mathematical analysis of the respective scattering
problem can be found in \cite{PRB03,PRE06}.

\subsubsection{Regular photonic band edge}

We start with the simplest case of a regular photonic band edge. There are
two different possibilities in this case, but none of them is associated
with the frozen mode regime. The first one relates to the trivial case where
none of the layers of the periodic structure displays an in-plane anisotropy
or gyrotropy. In this case, all eigenmodes are doubly degenerate with
respect to polarization. A detailed description of this case can be found in
the extensive literature on optics of stratified media \cite{Strat1,Strat3}.
Slightly different scenario occurs if some of the layers are anisotropic or
gyrotropic and, as a result, the polarization degeneracy is lifted. Just
below the band edge frequency $\omega _{g}$ in Fig. \ref{DRSP3}(a), the
transmitted field $\Psi _{T}\left( z\right) $ is a superposition (\ref%
{PsiT=pr+ev}) of one propagating and one evanescent Bloch modes. Due to the
boundary condition (\ref{BC}), the amplitude of the transmitted wave at $z=0$
is comparable to that of the incident wave. In the case of a generic
polarization of the incident light, the amplitudes of the propagating and
evanescent Bloch components at $z=0$ are also comparable to each other and
to the amplitude of the incident light%
\begin{equation}
\left\vert \Psi _{pr}\left( 0\right) \right\vert \sim \left\vert \Psi
_{ev}\left( 0\right) \right\vert \sim \left\vert \Psi _{I}\right\vert ,\text{
at }\ \omega \leq \omega _{g}.  \label{pr = ev = I}
\end{equation}%
As the distance $z$ from the surface increases, the evanescent component $%
\Psi _{ev}\left( z\right) $ decays rapidly, while the amplitude of the
propagating component remains constant. Eventually, at a certain distance
from the slab surface, the transmitted wave $\Psi _{T}\left( z\right) $
becomes very close to the propagating mode%
\begin{equation}
\Psi _{T}\left( z\right) \approx \Psi _{pr}\left( z\right) ,\text{ at }z\gg
L,\ \omega \leq \omega _{g}.  \label{PsiT=Psipr}
\end{equation}

The evanescent component $\Psi_{ev}$ of the transmitted wave does not
display any singularity at the band edge frequency $\omega_{g}$. The
propagating mode $\Psi_{pr}$ does develop a singularity associated with
vanishing group velocity at $\omega\rightarrow\omega_{g}-0$, but its
amplitude remains finite and comparable to that of the incident wave. At $%
\omega>\omega_{g}$, this propagating mode turns into another evanescent mode
in (\ref{PsiT=ev+ev}). The bottom line is that none of the Bloch components
of the transmitted wave develops a large amplitude in the vicinity of a
regular photonic band edge. There is no frozen mode regime in this trivial
case.

\subsubsection{Stationary inflection point}

\begin{figure}[tbph]
\scalebox{0.8}{\includegraphics[viewport=0 0 500 180,clip]{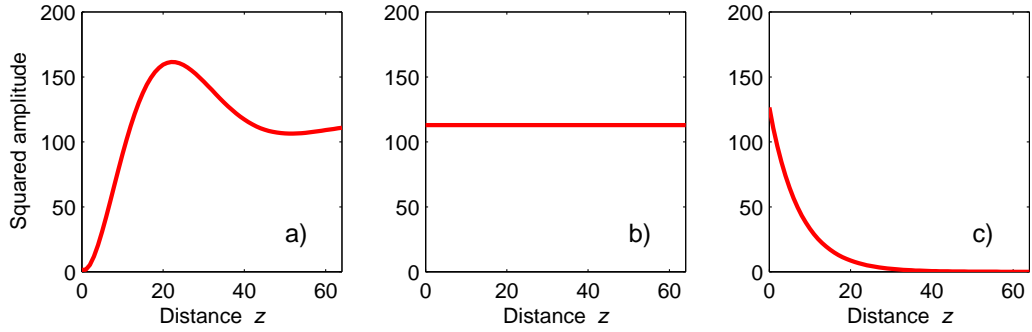}}
\caption{(Color online). Destructive interference of the propagating and
evanescent components of the transmitted wave inside semi-infinite photonic
crystal. The frequency is close but not equal to that of stationary
inflection point. (a) The squared modulus of the resulting transmitted field
-- its amplitude at $z=0$ is small enough to satisfy the boundary
conditions; (b) the squared modulus of the propagating contribution, which
is independent of $z$; (c) the squared modulus of the evanescent
contribution, which decays with the distance $z$. The amplitude of the
incident wave is unity. The distance $z$ from the surface is expressed in
units of $L$.}
\label{Amz_AAF}
\end{figure}

A completely different situation develops in the vicinity of a stationary
inflection point (\ref{SIP DR}). At $\omega \approx \omega _{0}$, the
transmitted wave $\Psi _{T}$ is a superposition (\ref{PsiT=pr+ev}) of one
propagating and one evanescent Bloch component. In contrast to the case of a
regular photonic band edge, in the vicinity of $\omega _{0}$ both Bloch
contributions to $\Psi _{T}$ develop strong singularity. Specifically, as
the frequency $\omega $ approaches $\omega _{0}$, both contributions grow
dramatically, while remaining nearly equal and opposite in sign at the slab
boundary \cite{PRB03}%
\begin{equation}
\Psi _{pr}\left( 0\right) \approx -\Psi _{ev}\left( 0\right) \propto
\left\vert \omega -\omega _{0}\right\vert ^{-1/3},\ \ \text{as }\omega
\rightarrow \omega _{0}.  \label{DI 3}
\end{equation}%
Due to the destructive interference (\ref{DI 3}), the resulting field%
\begin{equation*}
\Psi _{T}\left( 0\right) =\Psi _{pr}\left( 0\right) +\Psi _{ev}\left(
0\right)
\end{equation*}%
at the surface at $z=0$ is small enough to satisfy the boundary condition (%
\ref{BC}). As the distance $z$ from the slab boundary increases, the
destructive interference becomes less effective -- in part because the
evanescent contribution decays exponentially%
\begin{equation}
\Psi _{ev}\left( z\right) \approx \Psi _{ev}\left( 0\right) \exp \left(
-zk^{\prime \prime }\right) ,  \label{Psi_ev SIP}
\end{equation}%
while the amplitude of the propagating contribution remains constant and
very large. Eventually, the transmitted wave $\Psi _{T}\left( z\right) $
reaches its large saturation value corresponding to its propagating
component $\Psi _{pr}$, as seen in Fig. \ref{Amz_AAF}(a).

Note that the imaginary part $k^{\prime\prime}$ of the evanescent mode wave
number in (\ref{Psi_ev SIP}) also vanishes in the vicinity of stationary
inflection point%
\begin{equation}
k^{\prime\prime}\propto\left\vert \omega-\omega_{0}\right\vert ^{1/3}\text{,
as }\omega\rightarrow\omega_{0},  \label{Imk SIP}
\end{equation}
reducing the rate of decay of the evanescent contribution (\ref{Psi_ev SIP}%
). As a consequence, the resulting amplitude $\Psi_{T}\left( z\right) $ of
the transmitted wave reaches its large saturation value $\Psi_{pr}$ in (\ref%
{DI 3}) only at a certain distance $Z$ from the surface%
\begin{equation}
Z\propto1/k^{\prime\prime}\propto\left\vert \omega-\omega_{0}\right\vert
^{-1/3}.  \label{Z SIP}
\end{equation}
This characteristic distance diverges as the frequency approaches its
critical value $\omega_{0}$.

If the frequency of the incident wave is exactly equal to the frozen mode
frequency $\omega_{0}$, the transmitted wave $\Psi_{T}\left( z\right) $ does
not reduce to the sum (\ref{PsiT=pr+ev}) of propagating and evanescent
contributions, because at $\omega=\omega_{0}$, there is no evanescent
solutions to the Maxwell equations. Instead, $\Psi_{T}\left( z\right) $
corresponds to a non-Bloch Floquet eigenmode diverging linearly with $z$ 
\cite{PRB03}. 
\begin{equation}
\Psi_{T}\left( z\right) -\Psi_{T}\left( 0\right) \propto z\Psi _{0},\;\ 
\text{at}\;\omega=\omega_{0}.  \label{FL SIP}
\end{equation}
Such a solution is shown in Fig. \ref{AMn6w0}(c).

\subsubsection{Degenerate band edge}

While the situation with a regular photonic band edge appears trivial, the
case of a degenerate band edge (\ref{DBE DR}) proves to be quite different.
Just below the degenerate band edge frequency $\omega_{d}$ (inside the
transmission band), the transmitted field is a superposition (\ref%
{PsiT=pr+ev}) of one propagating and one evanescent components. Above $%
\omega_{d}$ (inside the band gap), the transmitted wave is a combination (%
\ref{PsiT=ev+ev}) of two evanescent components. In this respect, a regular
and a degenerate band edges are similar to each other. A crucial difference,
though, is that in the vicinity of a degenerate band edge, both Bloch
contributions to the transmitted wave diverge as $\omega$ approaches $%
\omega_{d}$, both above and below the band edge frequency. This constitutes
the frozen mode regime.

\begin{figure}[tbph]
\scalebox{0.8}{\includegraphics[viewport=0 0 500 180,clip]{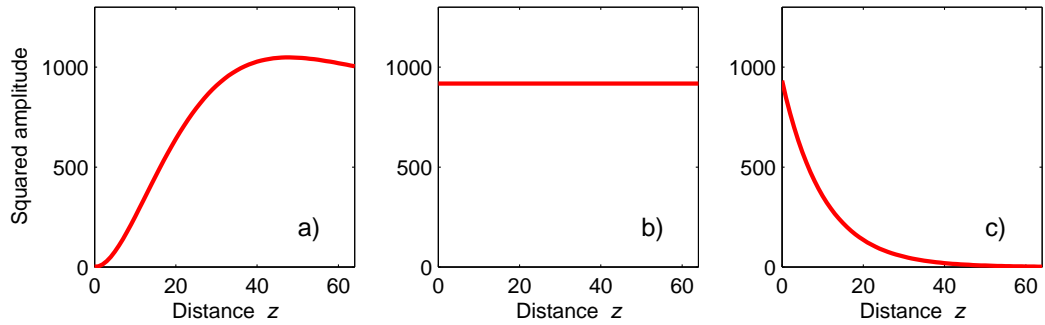}}
\caption{(Color online). Destructive interference of the two Bloch
components of the transmitted wave inside semi-infinite photonic crystal.
The frequency is $\protect\omega =\protect\omega _{d}-10^{-4}c/L$, which is
slightly below the degenerate band edge in Fig. \protect\ref{DRSP3}(c). (a)
The squared modulus of the resulting transmitted field -- its amplitude at $%
z=0$ is small enough to satisfy the boundary conditions (\protect\ref{BC});
(b) the squared modulus of the propagating contribution, which is
independent of $z$; (c) the squared modulus of the evanescent contribution,
which decays with the distance $z$. The amplitude of the incident wave is
unity.}
\label{Amn_B3}
\end{figure}

Let us start with the transmission band. As the frequency $\omega $
approaches $\omega _{d}-0$, both Bloch contributions in (\ref{PsiT=pr+ev})
grow sharply, while remaining nearly equal and opposite in sign at the
surface at $z$ = 0 \cite{PRE06}%
\begin{equation}
\Psi _{pr}\left( 0\right) \approx -\Psi _{ev}\left( 0\right) \propto
\left\vert \omega _{d}-\omega \right\vert ^{-1/4},\ \ \text{as }\omega
\rightarrow \omega _{d}-0.  \label{DI 4b}
\end{equation}%
The destructive interference (\ref{DI 4b}) ensures that the boundary
condition (\ref{BC}) can be satisfied, while both Bloch contributions to $%
\Psi _{T}\left( z\right) $ diverge. As the distance $z$ from the slab
boundary increases, the evanescent component $\Psi _{ev}\left( z\right) $
dies out%
\begin{equation}
\Psi _{ev}\left( z\right) \approx \Psi _{ev}\left( 0\right) \exp \left(
-zk^{\prime \prime }\right)  \label{Psi_ev DBE}
\end{equation}%
while the propagating component $\Psi _{pr}\left( z\right) $ remains
constant and very large. Eventually, as the distance $z$ further increases,
the transmitted wave $\Psi _{T}\left( z\right) $ reaches its large
saturation value corresponding to its propagating component $\Psi
_{pr}\left( z\right) $, as illustrated in Fig. \ref{Amn_B3}. Note that the
imaginary part $k^{\prime \prime }$ of the evanescent mode wave number also
vanishes in the vicinity of degenerate band edge%
\begin{equation}
k^{\prime \prime }\propto \left\vert \omega -\omega _{d}\right\vert ^{1/4}%
\text{, as }\omega \rightarrow \omega _{d},  \label{Imk DBE}
\end{equation}%
reducing the rate of decay of the evanescent contribution (\ref{Psi_ev DBE}%
). As a consequence, the resulting amplitude $\Psi _{T}\left( z\right) $ of
the transmitted wave reaches its large saturation value $\Psi _{pr}$ only at
a certain distance $Z$ from the surface%
\begin{equation}
Z\propto 1/k^{\prime \prime }\propto \left\vert \omega -\omega
_{d}\right\vert ^{-1/4}.  \label{Z DBE}
\end{equation}%
This characteristic distance increases as the frequency approaches its
critical value $\omega _{d}$, as illustrated in Fig. \ref{Amn6wd}(a) and (b).

If the frequency $\omega$ of the incident wave is exactly equal to $\omega
_{d}$, the transmitted wave $\Psi_{T}\left( z\right) $ does not reduce to
the sum of two Bloch contributions. Instead, it corresponds to a non-Bloch
Floquet eigenmode linearly diverging with $z$ 
\begin{equation}
\Psi_{T}\left( z\right) -\Psi_{T}\left( 0\right) \propto z\Psi _{d},\;\ 
\text{at}\;\omega=\omega_{d}.  \label{FL DBE}
\end{equation}
This situation is shown in Fig. \ref{Amn6wd}(c).

The above behavior appears to be very similar to that of the frozen mode
regime at a stationary inflection point, shown in Figs. \ref{AMn6w0} and \ref%
{Amz_AAF}. Yet, there is a crucial difference between the frozen mode regime
at a stationary inflection point and at a degenerate band edge. In the
immediate proximity of a degenerate band edge, the Pointing vector $S_{T}$
of the transmitted wave is infinitesimal, in spite of the diverging wave
amplitude. In other words, although the energy density $W_{T}\propto\left%
\vert \Psi_{T}\right\vert ^{2}$ of the frozen mode diverges as $\omega
\rightarrow\omega_{d}-0$, it does not grow fast enough to offset the
vanishing group velocity. As a consequence, the photonic crystal becomes
totally reflective at $\omega=\omega_{d}$. Of course, the total reflectivity
persists at $\omega>\omega_{d}$, where there is no propagating modes at all.
By contrast, in the case (\ref{FL SIP}) of a stationary inflection point,
the respective Pointing vector $S_{T}$ is finite and can be even close to
that of the incident wave, implying low reflectivity and nearly total
conversion of the incident wave energy into the frozen mode.

\begin{figure}[tbph]
\scalebox{0.8}{\includegraphics[viewport=0 0 500 180,clip]{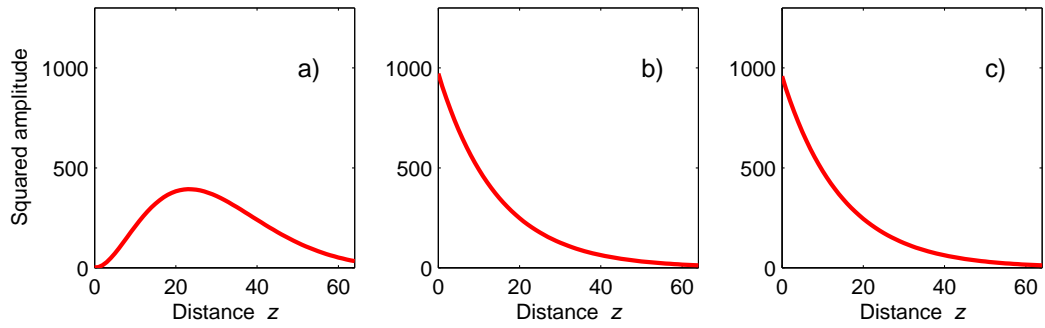}}
\caption{(Color online). Destructive interference of the two Bloch
components of the transmitted wave inside semi-infinite photonic crystal.
The frequency is $\protect\omega =\protect\omega _{d}+10^{-5}c/L$, which is
just above the degenerate band edge. (a) The squared modulus of the
resulting transmitted field -- its amplitude at $z=0$ is small enough to
satisfy the boundary conditions (\protect\ref{BC}); (b) and (c) the squared
moduli of the two evanescent contributions; both decay with the distance $z$%
. The amplitude of the incident wave is unity.}
\label{Amn_G3}
\end{figure}

The character of frozen mode regime is different when we approach the
degenerate band edge frequency from the band gap. In such a case, the
transmitted field $\Psi _{T}\left( z\right) $ is a superposition (\ref%
{PsiT=ev+ev}) of two evanescent components. As the frequency $\omega $
approaches $\omega _{d}$, both evanescent contributions grow sharply, while
remaining nearly equal and opposite in sign at the photonic crystal boundary%
\begin{equation}
\Psi _{ev1}\left( 0\right) \approx -\Psi _{ev2}\left( 0\right) \propto
\left\vert \omega _{d}-\omega \right\vert ^{-1/4},\ \ \text{as }\omega
\rightarrow \omega _{d}+0.  \label{DI 4g}
\end{equation}%
Again, the destructive interference (\ref{DI 4g}) ensures that the boundary
condition (\ref{BC}) can be satisfied, while both evanescent contributions
to $\Psi _{T}\left( z\right) $ diverge in accordance with (\ref{DI 4g}). As
the distance $z$ from the slab boundary increases, the destructive
interference of these two evanescent components is lifted and the resulting
field amplitude increases sharply, as seen in Fig. \ref{Amn_G3}(a). But
eventually, as the distance $z$ further increases, the transmitted wave $%
\Psi _{T}\left( z\right) $ completely decays, because both Bloch
contributions to $\Psi _{T}\left( z\right) $ are evanescent. The latter
constitutes the major difference between the frozen mode regime above and
below the DBE frequency $\omega _{d}$. The rate of the amplitude decay, as
well as the position of the maximum of the transmitted wave amplitude in
Figs. \ref{Amn6wd}(d -- f) and \ref{Amn_G3}(a) , are characterized by the
distance $Z$ in (\ref{Z DBE}).

\subsubsection{Physical reason for the growing wave amplitude}

If the frequency $\omega $ is close, but not equal, to that of a stationary
point (\ref{SP}) of the dispersion relation, the wave $\Psi _{T}\left(
z\right) $ transmitted to the semi-infinite periodic layered structure is a
superposition of two forward Bloch modes $\Psi _{1}\left( z\right) $ and $%
\Psi _{2}\left( z\right) $%
\begin{equation}
\Psi _{T}\left( z\right) =\Psi _{1}\left( z\right) +\Psi _{2}\left( z\right)
,~z\geq 0.  \label{Psi1+Psi2}
\end{equation}%
The two Bloch modes in (\ref{Psi1+Psi2}) can be one propagating and one
evanescent, as in (\ref{PsiT=pr+ev}), or they can be both evanescent, as in (%
\ref{PsiT=ev+ev}). In the vicinity of frozen mode regime, as the frequency
approaches its critical value of $\omega _{0}$ or $\omega _{d}$, the
four-dimensional vectors (\ref{Psi}) corresponding to each of the two Bloch
eigenmodes on the right-hand side of (\ref{Psi1+Psi2}) become nearly
parallel to each other 
\begin{equation}
\Psi _{1}\left( z\right) \approx \alpha \Psi _{2}\left( z\right) ,\ \ \text{%
as }\omega \rightarrow \omega _{s},  \label{Psi - Psi}
\end{equation}%
where $\alpha $ is a scalar, and $\omega _{s}$ is the frozen mode frequency $%
\omega _{0}$ or $\omega _{d}$. The asymptotic relation (\ref{Psi - Psi})
reflects a basic property of the transfer matrix of the periodic layered
structure at the frequency of either stationary point $\omega _{0}$ or $%
\omega _{d}$ (see [36-38]).

Let us show how the property (\ref{Psi - Psi}) of the Bloch eigenmodes in
the vicinity of $\omega _{0}$ or $\omega _{d}$ can lead to the frozen mode
regime. Indeed, at the photonic crystal boundary, the sum (\ref{Psi1+Psi2})
of two nearly parallel column vectors $\Psi _{1}\left( z\right) $ and $\Psi
_{2}\left( z\right) $ must match the boundary conditions (\ref{BC}) with the
incident and reflected waves $\Psi _{I}\left( 0\right) $ and $\Psi
_{R}\left( 0\right) $. If the incident wave polarization is general, we have
no reason to expect the column-vector%
\begin{equation}
\Psi \left( 0\right) =\Psi _{I}\left( 0\right) +\Psi _{R}\left( 0\right) 
\label{PsiI+PsiR}
\end{equation}%
at the photonic crystal interface to be nearly parallel to $\Psi _{1}\left(
0\right) $ and $\Psi _{2}\left( 0\right) $. But on the other hand, the
boundary condition (\ref{BC}) says that%
\begin{equation}
\Psi \left( 0\right) =\Psi _{1}\left( 0\right) +\Psi _{2}\left( 0\right) 
\label{Psi(0)}
\end{equation}%
Obviously, the only situation where $\Psi \left( 0\right) $ in (\ref%
{PsiI+PsiR}) and (\ref{Psi(0)}) is a general four-dimensional vector, while $%
\Psi _{1}\left( 0\right) $ and $\Psi _{2}\left( 0\right) $ on the right-hand
side of (\ref{Psi(0)}) are nearly parallel to each other is when%
\begin{equation}
\Psi _{1}\left( 0\right) \approx -\Psi _{2}\left( 0\right) ,\ \ \left\vert
\Psi _{1}\left( 0\right) \right\vert \approx \left\vert \Psi _{2}\left(
0\right) \right\vert \gg \left\vert \Psi \left( 0\right) \right\vert .
\label{Psi1= -Psi2}
\end{equation}%
The relation (\ref{Psi1= -Psi2}) explains how the huge diverging amplitude
of the frozen mode inside the photonic crystal can be reconciled with the
boundary conditions (\ref{BC}).

There is one exception, though. As we already stated in (\ref{Psi - Psi}),
in the vicinity of the frozen mode frequency, the two Bloch components $\Psi
_{1}\left( z\right) $ and $\Psi _{2}\left( z\right) $ of the transmitted
wave $\Psi _{T}\left( z\right) $ are nearly parallel to each other. For this
reason, if the polarization of the incident wave $\Psi _{I}$ is such that $%
\Psi \left( 0\right) $ in (\ref{Psi(0)}) is nearly parallel to one of the
Bloch eigenmodes $\Psi _{1}\left( 0\right) $ or $\Psi _{2}\left( 0\right) $,
it is also nearly parallel to the other one. So, all three column vectors $%
\Psi _{1}\left( 0\right) $, $\Psi _{2}\left( 0\right) $, and $\Psi \left(
0\right) $ are now parallel to each other. In this, and only this case, the
amplitude of the transmitted wave $\Psi _{T}\left( z\right) $ will be
comparable to that of the incident wave. There is no frozen mode regime for
the respective vanishingly small range of the incident wave polarization. A
particular case of the above situation is the regime of a single mode
excitation, where only one of the two Bloch components $\Psi _{1}\left(
z\right) $ or $\Psi _{2}\left( z\right) $ in (\ref{Psi1+Psi2}) contributes
to the transmitted wave $\Psi _{T}\left( z\right) $ \cite{PRE06}. Indeed, if
the transmitted wave is a single Bloch mode (either $\Psi _{1}\left(
z\right) $, or $\Psi _{2}\left( z\right) $),  the condition (\ref{Psi1=
-Psi2}) cannot be satisfied.

Finally, let us reiterate that in the limiting cases of $\omega =\omega _{0}$
or $\omega =\omega _{d}$, the transmitted wave $\Psi _{T}\left( z\right) $
corresponds to the non-Bloch Floquet eigenmode (\ref{FL SIP}) or (\ref{FL
DBE}), respectively. Either of them linearly diverges with $z$. The only
exception is when the incident wave has the unique (elliptic) polarization,
for which the transmitted wave $\Psi _{T}\left( z\right) $ is a propagating
Bloch eigenmode with zero group velocity and a limited amplitude, comparable
to that of the incident wave. Incident wave with any other polarization will
generate the frozen mode inside the periodic medium.

\section{Summary}

Although the existence of slow electromagnetic modes in photonic crystals is
quite obvious, the next question is whether and how such modes can be
excited by incident light. We have shown that the frozen mode regime
provides unique advantages in this respect. Generally, the possibility of
the frozen mode regime is determined by the character of the Bloch
dispersion relation, rather than the specific physical realization of the
periodic structure supporting such a dispersion relation. As soon as the
Bloch dispersion relation displays the proper singularity such as a
stationary inflection point (\ref{SIP DR}) or a degenerate band edge (\ref%
{DBE DR}), we have every reason to expect the occurrence of the frozen mode
regime at the respective frequency. In other words, the possibility of the
frozen mode regime is determined by the spectral properties of the periodic
structure, rather than by specific physical nature of the linear excitations.

If a periodic structure is relatively simple -- for instance, a stratified
medium with one dimensional periodicity -- its frequency spectrum may prove
to be too simple to support the proper spectral singularity and the frozen
mode regime. All our numerical examples relate to $1D$ photonic crystals,
where the existence of the frozen mode regime requires the presence of
misaligned birefringent layers (see the details in \cite{PRE06}). The more
complex the periodic structure is, the more likely it is capable of
supporting such a phenomenon. In photonic crystals with $3D$ periodicity,
there are no fundamental restrictions on the existence of the frozen mode
regime \cite{Hui08}. The same is true for the cases of modulated waveguides
and periodic arrays of coupled resonators.

Another important question is how robust the frozen mode regime is. For
instance, what happens if we introduce absorption or structural
imperfections. Of course, all these factors suppress the frozen mode
amplitude, but not as much as in the case of common Fabry-Perot or
transmission band edge resonances, where the entire photonic structure works
as a resonator. In addition, since in the vicinity of an inflection point
the dispersion term $v_{g}^{\prime }\left( k\right) =\omega ^{\prime \prime
}\left( k\right) $ vanishes, one can achieve wide-bandwidth and
dispersion-free propagation of light \cite{PCCW B-08R}.

Finally, it is possible to combine the frozen mode regime and a common slow
wave resonance, in which case we have the phenomenon called the giant
transmission resonance \cite{PRE05,PRA07}. The Q-factor associated with such
a resonance can be by two orders of magnitude higher, compared to that of
the regular Fabry-Perot resonance in the same or similar periodic
structure.\bigskip

\textbf{Acknowledgment and Disclaimer:} Effort of A. Figotin and I.
Vitebskiy is sponsored by the Air Force Office of Scientific Research, Air
Force Materials Command, USAF, under grant number FA9550-04-1-0359.

\end{document}